# Role of thermal vibrations in phase transitions

T.R.S. Prasanna

Department of Metallurgical Engineering and Materials Science

Indian Institute of Technology, Bombay

Mumbai, 400076 India

All theoretical models (Heisenberg, Ising etc.) assume a negligible role for thermal vibrations in alloy and magnetic phase transitions. Analysis of diffraction data *conclusively proves* that this assumption is incorrect. A simple criterion emerges that theoretical models can ignore the role of thermal vibrations only if the Debye-Waller Factor is ignored in the analysis of diffraction data. Diffraction data constrain all theoretical models to incorporate the role of thermal vibrations. This conclusion is also supported by other experimental results, the effect of thermal vibrations on ordering energy that is of the same order of magnitude as ordering energy and an isotope effect on magnetic phase transitions. An electron-phonon interaction (EPI) formalism that incorporates the Debye-Waller Factor in electronic structure calculations already exists and must be adopted for a correct understanding of phase transitions as it can account for all the different experimental results mentioned above. The discrepancy between experimental and theoretical ordering energy in $Ni_3V$ is direct evidence for the role of thermal vibrations in altering ordering energy. The inter-nuclear potential energy term converges if zero point vibrations are incorporated and this method can replace the Ewald sum method. The three dimensional Ising model cannot represent order-disorder transition in beta brass, CuZn. An isotope effect is predicted for magnetic phase transitions if the transition temperature is below Debye temperature. The long range order parameter obtained from diffraction data can only be compared with predictions of models that incorporate the role of thermal vibrations and not otherwise.

# 1. Introduction

Order-disorder transitions have been the subject of intense study over decades. The first model for order-disorder transition in alloys was the Bragg-Williams (BW) model [1-3] that proposed the existence of a long-range order parameter. The deficiencies [1-3] of this model include a higher transition temperature, $T_c$, and failure to account for specific heat above $T_c$ due to the neglect of fluctuations. The cluster variation method [4] is a generalization of this approach that can be applied to the generalized Ising model where next nearest neighbor interactions are also incorporated. In parallel, models were proposed for magnetic ordering, the most influential among them being the Ising model. The Ising model is the simplest of nearest neighbor models that consider the cooperative interaction between neighboring spins. The exact solution to the two-dimensional (2D) Ising model [5] continues to be a landmark as the solution for the 3D Ising model has proved elusive. Recently, the author has described a method to determine the form of the exact partition function for the Ising model [6]. The BW model is equivalent to a mean-field solution of the Ising model.

The study of phase transitions has largely relied on models as first principles theories have proved difficult, especially in the study of magnetism. The prediction of phase diagrams [7-10] from first principles is a related problem of great interest and one where considerable progress has been made. The development of density functional theory [11-13] is a landmark that has contributed significantly to this effort.

In most studies of phase transitions, a model is proposed where an interaction parameter, $J_{ij}$, is assumed between neighboring elements. The Heisenberg model can be considered to be a prototype model and is represented as

$$H = -\sum J_{ij}\, \mathbf{S}_i \cdot \mathbf{S}_j \tag{1}$$

The Ising model is a special case of this model. In this paper we will consider Eq.1 to be representative of most models of alloy and magnetic phase transitions. All phase transitions occur at finite temperatures where thermal vibrations are present. However, the interaction parameter, $J_{ij}$, in Eq.1 is for a static lattice. The implicit assumption in Eq.1 is that the role of thermal vibrations in phase transitions is negligible. According to this implicit assumption, whether the role of thermal vibrations is incorporated or ignored makes little difference to the understanding of phase transitions. One consequence of Eq.1 is that the ordering energy is assumed to be constant in all models of phase transitions. In this paper we examine the validity of the assumption that thermal vibrations play a negligible role and conclusively establish that this assumption is incorrect. That is, the interaction parameter, $J_{ij}$, is a function of temperature due to the role of thermal vibrations and must be represented as $J_{ij}(T)$ in all theories and models of phase transitions. In addition, new insights into phase transitions are obtained from the analysis presented in this study.

**2. Preliminary Analysis**

Before embarking on a detailed analysis, we present a preliminary analysis of the role of thermal vibrations on phase transitions.

The density functional theorem [11-13] states that for a given external potential V(**r**), the ground state energy is a functional of the charge density, E[$\rho(r)$], and is minimized for the correct charge density. Under static lattice conditions, the ordering energy can be symbolically represented as $\Delta E^{ord} = E\left[\rho^{ord}(r)\right] - E\left[(\rho^{dis}(r)\right]$. However, it is routinely observed in x-ray diffraction experiments that thermal vibrations alter the charge density. It follows that the ground state energies of both the ordered and disordered phases will be different at finite temperatures from their static lattice values. Hence, the ordering energy is a function of temperature due to the role of thermal vibrations.

The above conclusion is also supported by the following physical argument. At finite temperatures, various atomic configurations are accessed due to thermal vibrations and the electron distribution is in instantaneous equilibrium with them. The electron distributions in equilibrium with various configurations are of higher energy compared to the static lattice value. Clearly, the ensemble average over these configurations will result in higher energy when compared to the 0 K value. This result is valid for both the ordered and disordered state. At 0 K, the difference in energy between them is the conventional ordering energy. At finite temperatures, ordering energy is the difference in the ensemble average energies of the ordered and disordered state. It would be an extraordinary coincidence if this difference were to be *exactly* equal to the difference at 0 K and cannot be assumed without justification. Hence, the correct conclusion is that the differences in ensemble average energies between the ordered and disordered state at high temperatures are different from the 0 K value. Hence, *ordering energy is a function of temperature*.

In an entirely different approach, the role of thermal vibrations on properties of materials is discussed in Ref.14. The basic premise is that thermal vibrations cause renormalization of physical properties due to which all properties are functions of temperature. Ref.14 states that "*the unrenormalized quantity $Q_0$ is a theoretical construct which does not exist in nature*". From the discussion in Ref.14, it is clear that the <u>theoretically correct representation of the interaction parameter is $J_{ij}(T)$</u> and the representation in most models as $J_{ij}$ is an approximation. The only question that needs to be addressed is how justified is the approximation or how significant is the role of thermal vibrations in phase transitions.

One approach is to perform electronic structure calculations for specific systems under static lattice conditions and also at finite temperatures, e.g., at transition temperatures, $T_c$, by incorporating the role of thermal vibrations. Such a study is yet to be performed for any alloy that undergoes an order-disorder transition. The limitation of this approach is that while it may establish the importance (or lack of it) of thermal vibrations in a specific system or a few specific systems, it would be a gross exaggeration to reach universal conclusions that are applicable to all theories and models of phase transitions. The computational approach is particularly well suited to study the details of a specific system and not suited to reach universal conclusions on phase transitions.

Another approach is to analyze experimental data that are universal to all systems that exhibit phase transitions.

The first experimental observation that we consider is that thermal vibrations alter electron energies. As stated in Ref.14, "*Electronic quasiparticle energies are renormalized by lattice vibrations giving a T dependence, a zero-point shift and an isotope dependence*". The temperature dependence is in the range of 2-4 $k_BT$ per degree [14-16]. Even by taking the lower value, thermal vibrations shift electron energies by ~ 0.2 meV/K in metals and semiconductors [15]. This implies shifts in electron energies of ~ $10^2$ meV near transition temperatures (~ 1000 K) for both ordered and disordered phases. *Their difference would be of the same order of magnitude as the ordering energy (10-100 meV) [8] and hence, the role of thermal vibrations cannot be ignored.*

A similar comment is made in Ref. 16 with regard to the importance of thermal vibrations in change of phase, though not phase transitions. Since the energy associated with order-disorder transitions are usually lower than that associated with phase change, it is clear that the role of thermal vibrations will be significant and cannot be ignored.

The analysis presented so far is based on extending existing ideas to phase transitions. This is sufficient to establish that the assumption that thermal vibrations play a negligible role in phase transitions is highly suspect and unjustified. That is, this assumption must be substantiated before it can be accepted as correct. However, it does not establish that this assumption is definitively incorrect. We present new analysis below that conclusively establishes that this assumption is incorrect.

## 3. Additional Analysis

The next experimental observation that we consider are diffraction data. X-ray, electron and neutron diffraction experiments are the most common method to obtain the state of (partial) order at any temperature. A common observation that is universal to all alloy and magnetic phase transitions is the emergence of superlattice lines upon ordering in diffraction experiments. Hence, the analysis of superlattice lines will have universal implications applicable to all theories and models of phase transitions.

In the analysis of x-ray, electron and neutron diffraction data to determine the state of partial order at any temperature, corrections for thermal vibrations through the Debye-Waller factor is essential [17-23]. Ignoring other factors, the intensity of a superlattice line, **G,** at any temperature is given by [17-23]

$$I_G(T) = I_G \, \eta^2 \, e^{-2M(G,T)} \tag{2}$$

where η is the order parameter, M is the Debye-Waller factor (DWF) and $I_G$ is a constant that is the intensity of the superlattice line at T = 0 K where the order parameter η = 1. The order parameter is obtained from Eq.2 as

$$\eta = \left[ I_G(T)/I_G \, e^{-2M(G,T)} \right]^{1/2} \tag{3}$$

Eq.3 can be used to examine the implicit assumption of all theoretical models that the role of thermal vibrations can be ignored as it makes little difference to the understanding of phase transitions whether the role of thermal vibrations is incorporated or ignored. To be consistent with this assumption, order parameter must be obtained from diffraction data using Eq.3 with the DWF being ignored, i.e. with M = 0. In this case, the order parameter is given by $\eta = [I_G(T)/I_G]^{1/2}$. One consequence is that *different superlattice lines, G, will result in different order parameters*. In addition, the correct order parameter, obtained by incorporating the DWF lies *outside the range of values* for the order parameter obtained under the assumption M = 0. Since different superlattice lines result in different order parameters, it is *not possible to speak of a state of order if the role of thermal vibrations is ignored* in the analysis of diffraction data. This is the reason that the role of thermal vibrations is <u>always</u> incorporated in the analysis of diffraction data. While the assumption of theoretical models is that the role of thermal vibrations can be ignored as incorporating or ignoring them has little consequence on phase transitions, *<u>diffraction data clearly show that the role of thermal vibrations must be incorporated and ignoring them has unacceptable consequences</u>*. Since experiments are the final arbiters of the correctness of theories, analysis of diffraction data <u>conclusively prove</u> that the assumption of all theoretical models of phase transitions that thermal vibrations play a negligible role is incorrect and hence, all such models must be corrected to incorporate the role of thermal vibrations.

As an illustration of the above analysis, we consider the x-ray [18] and neutron diffraction [19] data for beta brass, β-CuZn, which is usually described by the three-dimensional Ising model. Ref. 18, 19 describe in great detail the corrections necessary to extract the order parameter from diffraction data. The authors of Ref.18 state *"It (the combined temperature factor) is the largest of the three correction factors. Its dominant contribution comes from the mean Debye-Waller factor ratio…"*. Table 1 of Ref.19 presents all the relevant data in tabular form. The order parameter is listed in the last column and is obtained from an expression similar to Eq.3 except that η is normalized to 1 at room temperature and not 0 K. From Table 1 of Ref.19, it is also possible to calculate the order parameter by ignoring the role of thermal vibrations, i.e. by setting $X_{DW} = 1$ at all temperatures in Table 1 of Ref.19. For example, at reduced temperature *t=0.017*, the correct order parameter from both superlattice lines, (100) and (111), is 0.48. If the role of thermal vibrations is ignored, then the order parameter from (100) superlattice line is 0.46 and from the (111) superlattice line is 0.41. We see that different superlattice lines give different order parameters and the correct value, 0.48, lies outside the range of values, in this case 0.46 and 0.41. Similar discrepancies can be calculated for other temperatures from the data in Table 1 of Ref.19. Clearly, the role of thermal vibrations cannot be ignored in the analysis of diffraction data. This explains why the role of DWF is <u>always</u> incorporated in the analysis of diffraction data.

The diffraction data for beta brass, β-CuZn, clearly show that the role of thermal vibrations must be incorporated to reach meaningful conclusions. But the 3D Ising model neglects the role of thermal vibrations. This contradiction is seen starkly in Fig. 3 of Ref.19 where the order parameter obtained from x-ray and neutron diffraction data are compared with the predictions of the compressible and rigid Ising models. Since the former incorporate the role of thermal vibrations and the latter ignore the role of thermal vibrations, clearly the comparison in Fig.3 of Ref.19 is invalid. For the comparison to be valid, both the analysis of diffraction data and theoretical models must either ignore the role of thermal vibrations or incorporate them. Since the former leads to unacceptable consequences when analyzing diffraction data, only the latter possibility is valid. That is, the order parameter obtained from diffraction data for beta brass, β-CuZn, must only be compared with predictions of those models that incorporate the role of thermal vibrations. Hence, the 3 D Ising model cannot represent phase transition in beta brass, β-CuZn.

We have selected beta brass, β-CuZn, for illustration as the detailed description of the treatment of x-ray and neutron diffraction data to determine the order parameter exists in published literature [18,19]. It is clear that similar care would be employed in the analysis of diffraction data for other systems as well though the details may not be published in literature. *Because analysis of diffraction data <u>always</u> incorporates the role of thermal vibrations, no theoretical model that ignores the role of thermal vibrations can be compared with the results of diffraction data.*

As seen above, analysis of diffraction data conclusively proves that the role of thermal vibrations must be incorporated in all theoretical models of phase transitions. The role of thermal vibrations enters analysis of diffraction data through the Debye-Waller factor (DWF). This is discussed in detail in standard references on x-ray diffraction theory [24-27]. In the harmonic approximation, with the assumption of rigid ions or pseudo atoms, the effect of thermal vibrations is to change the Fourier Transform of the atomic charge density, atomic scattering factor, $f_j$ to $f_j e^{-M_j}$ where $M_j$ is the DWF [24-27]. In x-ray diffraction, the structure factor, $F_G$, is defined as the Fourier transform of the thermal-average charge density, $\overline{\rho(r)}$, and given by [24]

$$F_G = \int \overline{\rho(r)} e^{2\pi i G \cdot r} d^3 r \qquad (4)$$

and the intensity of a Bragg reflection is proportional to $|F_G|^2$. The intensity measured in a x-ray diffraction experiment is due to elastic scattering from time varying electron distributions that are in instantaneous equilibrium with various atomic configurations due to thermal vibrations [24,25]. However, for crystals the difference between the two intensities is negligible and $\Delta I/I_{Bragg} \sim O(N^{-1})$ [24]. Therefore, *"To an extremely good approximation, the scattering averaged over the instantaneous distributions is equivalent to the scattering of the time-averaged distribution of the scattering matter"* [26]. Hence, scattering of x-rays by crystals can be interpreted as resulting from a thermal-average electron density, $\overline{\rho(r)}$, in a static lattice [24-26]. Similar arguments are applicable in analysis of electron diffraction data [20].

From the theory of x-ray and electron diffraction, it is clear that the role of thermal vibrations is to alter the Fourier transform of the charge density or the potential by the DWF. More correctly, this is applicable only to the charge density or potential that remains rigid or is unchanged by thermal vibrations. This suggests that, at finite temperatures, this correction would be applicable to the potential due to the nucleus and the core electrons.

It is of great significance that a formalism for electronic structure calculations where the role of thermal vibrations is incorporated in an similar manner *already exists* [15,28]. This formalism, which we will call the electron-phonon interaction (EPI) formalism shows that the role of thermal vibrations result in two correction terms – one in which the role of DWF must be incorporated in electronic structure calculations (ES(DWF)) and a self energy (SE) term. The correct approach [15] at high temperatures is to first perform electronic structure calculations using the ES(DWF) formalism and use the results obtained to calculate the self-energy corrections. This formalism is the first recourse to account for the changes in valence electron properties in semiconductors and metals [29-35]. Frequently in practice, only the DWF is incorporated through the ES(DWF) formalism and the results are compared with experiments [29-35]. The ES(DWF) formalism is of the greatest significance in any theoretical understanding of phase transitions as it is <u>completely compatible</u> with theory underlying diffraction techniques, which are the primary experimental tools to study phase transitions.

Using the ES(DWF) formalism, it is possible to analyze the changes in various components of the total ordering energy with temperature. The total energy can be represented as

$$E_{tot} = E_{kin} + E_{n-n} + E_{c-c} + E_{c-n} + E_{v-n} + E_{v-c} + E_{v-v} + E_{xc} \qquad (5)$$

While they are usually evaluated in real space, Fourier (or reciprocal space) methods have also been developed [36,37]. The latter incorporate the periodicity of the lattice and various terms are expressed as sums over reciprocal lattice vectors [36,37]. Since ES(DWF) also changes the potential in reciprocal space, it is compatible with the method of Ref.36,37 though corrections have to be made for the role of thermal vibrations.

We begin with an analysis of the nuclear-nuclear energy term. From standard textbooks in electron microscopy [20], the structure factor due to nuclear charges, $Z_G$, is given by

$$Z_G = \sum_{j=1}^{N} Z_j e^{-M_j} e^{2\pi i \mathbf{G} \cdot \mathbf{r}_j} \qquad (6)$$

where $Z_j$ and $M_j$ are the charge and DWF for nucleus $j$, $\mathbf{r}_j$ its position in the unit cell and the summation is over all (N) atoms in the unit cell. $\mathbf{G}$ is a reciprocal lattice vector whose magnitude is given by $|\mathbf{G}|^2 = 1/d_{hkl}^2$ where $d_{hkl}$ is the interplanar spacing [17]. In the above expression, since the nuclear charge distribution can be considered to be 'rigid' with respect to thermal vibrations, DWF accounts for their role to a very good approximation. From the Poisson equation it follows that the Fourier transform of the potential due to nuclear charges is given by

$$V_{\mathbf{G}}^n = Z_{\mathbf{G}}/\mathbf{G}^2 \tag{7}$$

*which is temperature dependent* and similar to those found in electron microscopy textbooks [20]. This is the nuclear potential that must be used in the ES(DWF) formalism. The nuclear-nuclear repulsion energy, $E_{n\text{-}n}$, is given by

$$E_{n-n} = \sum_{\mathbf{G} \neq 0} \frac{|Z_{\mathbf{G}}|^2 e^{-2M}}{\mathbf{G}^2} \tag{8}$$

The same DWF (M) has been assumed for all atoms in the unit cell in Eq.8. The $\mathbf{G} = 0$ term is not considered as the average charge and potential in a unit cell is zero [38].

Before, we analyze the change in $E_{n\text{-}n}$ due to ordering, we discuss the Ewald summation method of evaluating $E_{n\text{-}n}$. It is well known [38,39] that the expression for $E_{n\text{-}n}$ does not converge e.g. Eq.F13 of Ref.38 (appendix F, p-640) and artificial parameters are introduced to ensure convergence. This is because this term is usually represented in the static approximation which ignores the role of thermal vibrations. However, even at 0 K, zero point vibrations are present and their role must be incorporated. *Hence, Eq.8 is the correct representation of $E_{n\text{-}n}$*. It is readily seen that with M = 0, Eq.8 reduces to Eq.F13 of Ref.38. The low temperature approximation to DWF, $M_L$, must be used in Eq.8 and is given by [38] $M_L = 3h^2|\mathbf{G}|^2/8mk\Theta_D$ where $|\mathbf{G}|^2 = 1/d_{hkl}^2$. It is readily seen that Eq.8 converges due to the presence of an exponential term that decays as $e^{-|\mathbf{G}|^2}$ ignoring constants. Thus the problem of convergence of $E_{n-n}$ is overcome if zero point vibrations are incorporated. Since Eq.8 represents the physics correctly and does not require any artificial parameters, it should replace the Ewald sum technique in evaluating $E_{n-n}$. An important advantage of this method is that it allows $E_{n-n}$ to be readily evaluated at *any temperature* if the DWF or mean-square displacements are known.

We next consider the change in the nuclear-nuclear energy term upon ordering. There is no change in structure factors, $Z_G$, upon ordering as the nuclear charge distribution remains "frozen". In the disordered phase, the summation in Eq.8 is only over fundamental lines, $\mathbf{G}_f$. In the ordered phase, the summation in Eq.8 is over fundamental lines, $\mathbf{G}_f$, and superlattice lines, $\mathbf{G}_s$. Comparison of the two shows that the change in nuclear-nuclear energy upon ordering, $\Delta E_{n-n}^{ord}$, *is stored only in superlattice wavevectors* and is given by

$$\Delta E_{n-n}^{ord} = \sum_{G_s} \frac{|Z_G|^2 e^{-2M}}{\mathbf{G}^2} \qquad (9)$$

At finite temperatures, the high temperature form, $M_H$, given by [17,38]

$$M_H = \frac{3h^2|\mathbf{G}|^2 T}{2mk\Theta_D^2} \qquad (10)$$

must be used. We stress that in both Eq.8 and Eq.9, we have approximated the DWF for different atoms, represented as $M_j$ in Eq.6, to be the same ($M$). This has been done to highlight explicitly the role of thermal vibrations on the $E_{n-n}$ energy term. More correctly, this energy term is represented as $E_{n-n} = |Z_G|^2 / G^2$ where $Z_G$ is given by Eq.6. It is readily seen that even in this more correct form, $E_{n-n}$ will converge due to the $M_j$ term.

We consider next the core electron- core electron energy term, $E_{e-e}$, The core electron density is frequently assumed to be 'rigid'. In this case, the effect of thermal vibrations can be incorporated by a DWF correction to the structure factor. The core electron – core electron interaction energy, $E_{c-c}$, is given in terms of Fourier components by

$$E_{c-c} = \sum_{G \neq 0} \frac{|F_G^c|^2 e^{-2M}}{G^2} \qquad (11)$$

assuming same DWF (M) for all atoms. $F_G^c$ is also given by

$$F_G^c = \sum_{j=1}^{N} f_j^c \, e^{-M_j} \, e^{2\pi i G \cdot r_j} \qquad (12)$$

where $f_j^c$ is the atomic scattering factor of core electrons and is the Fourier Transform of the core atomic charge density of atom $j$. The potential due to core electrons is

$$V_G^c = F_G^c / G^2 \qquad (13)$$

and *is temperature dependent*. This is the potential due to core electrons that must be used in the ES(DWF) formalism. Following arguments as in case of $E_{n-n}$, the entire change in core electron-electron energy upon ordering, $\Delta E_{c-c}^{ord}$, is given by

$$\Delta E_{c-c}^{ord} = \sum_{G_s} \frac{|F_G^c|^2 e^{-2M}}{G^2} \qquad (14)$$

and *stored only in superlattice lines*. We also see from Eq.14 that the energy is a function of temperature due to the DWF factor. From the above discussion, it follows that the ordering energy between the nucleus and core electrons, $\Delta E^{ord}_{n-c}$, is also a function of temperature of the type $e^{-2M}$ and *stored only in superlattice lines*. Therefore, all core energy components of the total ordering energy, $\Delta E^{ord}_{n-n}$, $\Delta E^{ord}_{c-c}$, $\Delta E^{ord}_{n-c}$, *are stored only in superlattice lines* and are functions of temperature.

The core energy terms have been obtained from diffraction theory [20] according to which the core potentials are given by Eq.7 and Eq.13. But Eq.7 and Eq.13 are also the core potentials used in the ES(DWF) formalism [15,29-35]. *Thus, diffraction theory is completely compatible with ES(DWF) formalism*. Therefore, it is possible to use Eq.7 and Eq.13 to analyze other energy terms that contribute to total energy. Since the core lattice potentials, Eq.7 and Eq.13, are temperature dependent, it follows that the valence electron distribution and energies will also be temperature dependent. It is readily seen that $\Delta E^{ord}_{v-n}$ and $\Delta E^{ord}_{v-c}$ will contain *at least* a temperature dependence of the type $e^{-M}$ that arises from the nuclear charge or core electron structure factors. It is clear that kinetic energy, valence electron-electron energy, exchange and correlation energy contributions to ordering energy will also be temperature dependent and these must be determined from electronic structure calculations. *Thus, all energy terms that contribute to the total ordering energy are temperature dependent.*

It is also possible to conclude that thermal vibrations *reduce* the ordering energy at finite temperatures from the static lattice values. The core lattice potentials, Eq.7 and Eq.13, of ordered and disordered phases differ only in the contributions from superlattice wavevectors in the former. The difference between the core potentials decreases at high temperatures due to the DWF. Since the core lattice potentials are "closer" at high temperatures than at 0 K, it follows that the total energies will also be "closer" at high temperatures than at 0 K. This strongly suggests that ordering energy will be less at high temperatures than at 0 K.

The assumption of all theoretical models that the ordering energy is independent of temperature implies that all components of the ordering energy must be independent of temperature. This implies that in electronic structure calculations the core potentials are independent of temperature. That is, in the ES(DWF) formalism, the DWF dependence of the core potentials in Eq.7 and Eq.13 must be ignored. But, since Eq.7 and Eq.13 are the core potentials used in diffraction theory as well, it follows that DWF must also be ignored in the analysis of diffraction data. This conclusion is possible due to the compatibility of ES(DWF) formalism with diffraction theory, which allows a simple criterion to test the validity of the assumption. That is, *ordering energy can be assumed to be independent of temperature only if the role of DWF can be ignored in the analysis of diffraction data.* Conversely, since DWF is <u>always</u> incorporated in the analysis of diffraction data, ordering energy must be represented as a function of temperature. Currently, standard references in alloy theory [7-9] assume that ordering energy is independent of temperature which can now be seen to be incorrect.

**4. Evidence from specific systems**

We next consider evidence for a temperature dependent ordering energy from different systems.

We consider beta-brass, $\beta$-CuZn, which exhibits an order-disorder transition [18,19] at 741 K. It has a Debye temperature, $\Theta_D = 263$ K [40] and lattice parameter of a = 2.95 Å. Instead of calculating the DWF, 2M, for which temperature needs to be specified, we write as $2M = \alpha T$ where the high temperature DWF (Eq.10) has been used. We see that $\alpha$ is independent of temperature. Its values for different superlattice lines are $\alpha_{100} = 1.5 \cdot 10^{-4}$ $K^{-1}$ for (100), $\alpha_{111} = 4.5 \cdot 10^{-4}$ $K^{-1}$ for (111) and $\alpha_{300} = 1.3 \cdot 10^{-3}$ $K^{-1}$ for superlattice line (300). Usually, total energy is evaluated for a static lattice (M = 0) in electronic structure calculations. The change in $\Delta E_{n-n}^{ord}$ at 725 K and 0 K (static lattice) can be determined from Eq.9. For $\beta$-CuZn, the inter-nuclear energy stored in different superlattice lines are reduced substantially. At 725 K, $\Delta E_{n-n}^{ord}(100)$, $\Delta E_{n-n}^{ord}(111)$ and $\Delta E_{n-n}^{ord}(300)$ are reduced to 89%, 72% and 37% of their respective 0 K values. Similar reductions would also occur for $\Delta E_{c-c}^{ord}$ and $\Delta E_{n-c}^{ord}$ as they have identical temperature dependence. These large changes in core energy contributions are alone sufficient to show that the 3D Ising model is incorrect as it does not contain a temperature dependence of ordering energy. Also, the DWF are different in the ordered and disorder phases of $\beta$-CuZn [18,19] which will contribute additionally to the difference in ordering energy at high temperatures compared to the 0 K value.

The simplest model of phase transitions is the Bragg-Williams model and the ordering energy can be made temperature dependent in the Modified Bragg-Williams model as

$$-\Delta E^{ord}(T) = E_0 \, e^{-\alpha_m T} \tag{15}$$

where $E_0$ is the total disordering energy at 0 K in the BW model. The configuration entropy remains same in BW and MBW models with the assumption of random distribution. The critical temperature, $T_c$, is given by

$$T_c = \frac{2 E_0 \, e^{-\alpha_m T_c}}{R} = T_c^{mft} \, e^{-\alpha_m T_c} \tag{16}$$

In BW model, the (mean field) critical temperature is given by $R\,T_c^{mft} = 2\,E_0$. The value of parameter $\alpha_m$ must be representative of thermal vibrations (~$10^{-3}$-$10^{-4}$ K$^{-1}$).

For beta-brass, β-CuZn, the experimental $T_c$ is 741 K. Assuming a nearest neighbor Ising model [18,19] with 8 neighbors, the mean field (BW) critical temperature is given by the relation $T_c/T_c^{mft} = 0.79385$ [41] to be $T_c^{mft}$ = 933.4 K. Assuming that $T_c$ in MBW model is the experimental value, 741 K, gives $\alpha_m$ to be 3.11 $10^{-4}$ K$^{-1}$. This is within the range set by the lowest and highest observed superlattice lines as $\alpha_{100}$ is 1.5 $10^{-4}$ K$^{-1}$ and $\alpha_{300}$ is 1.3 $10^{-3}$ K$^{-1}$ as discussed above. Thus, a mean $\alpha_m$ that is representative of thermal vibrations is sufficient to lower $T_c$ from the mean-field value of 933 K to the observed value of 741 K. It is stressed that the above discussion is meant to highlight the role of thermal vibrations and is not meant to be accurate. This is seen from the expression for the ordering energy, $E_0 \, e^{-\alpha_m T}$, where the exponential form has been adapted from the inter-nuclear and core electron energy terms, Eq.9 and Eq.14, as the temperature dependence of other energy terms is unknown. However, it is readily seen that $T_c$ will be altered substantially if even a fraction of the total ordering energy has exponential temperature dependence. Thus, even a simple MBW model shows that the role of thermal vibrations cannot be ignored.

Another example where the effect of thermal vibrations on ordering energy is very significant is $Ni_3V$. In this alloy [42-44], the experimentally observed ordering energy (10 meV/atom) at 1400 K and the theoretically determined ordering energy (100 meV/atom) at 0 K differ by almost a factor of 10. Electronic excitations and spin polarizations [42,44] do not fully account for this discrepancy. An explanation for the discrepancy relies on truncated cluster expansions [45] though there is no consensus in the alloy theory community [46]. There is no suggestion in literature till date that this discrepancy is due to the role of thermal vibrations in altering ordering energy. One possible reason could be the lack of awareness of the ES(DWF) or EPI formalism as evidenced from the absence of any mention of Ref.14-16 and Ref.29-35 in standard references on alloy theory [7-9] or in studies on $Ni_3V$ [42-46].

Based on the present study, the most probable reason for this discrepancy is the neglect of thermal vibrations that are present at 1400 K but absent in a static lattice calculation. Since thermal vibrations are the most significant difference at 1400 K over static lattice conditions, its effect on ordering energy must be considered first. If not, the role of DWF must be ignored in the analysis of diffraction data. Other explanations become necessary only if the role of thermal vibrations cannot fully account for changes in ordering energy.

Hence, Ni$_3$V *provides direct experimental evidence for the conclusion in this study that the ordering energy is a function of temperature due to the role of thermal vibrations.* Performing electronic structure calculations using the EPI (ES(DWF) + SE) formalism would allow the determination of the contribution of thermal vibrations of ES(DWF) and SE corrections. Only if these two corrections terms cannot account for all the observed changes do other explanations become important.

As mentioned earlier, no first principles electronic structure calculation exists till date where the role of thermal vibrations has been incorporated to study changes in ordering energy with temperature. However, such a calculation exists where the energy differences between two phases have been studied. *Ab Initio* Molecular dynamics (AIMD) simulations [47] reveal that the lattice stability of bcc Mo over fcc Mo is reduced to 0.2 eV at 3200 K from 0.4 eV at 0K. This shows that the role of thermal vibrations on lattice stability is of the same order of magnitude as lattice stability itself. However, as mentioned earlier, such a prediction was made almost 30 years ago in Ref.16. Importantly, the prediction in Ref.16 is not based on any particular formalism, EPI or AIMD, but based on the universal experimental observation that changes in total energies due to thermal vibrations are of the same order of magnitude as energy differences between phases. The results of Ref.47 are thus a computational confirmation of the prediction [16] made 30 years ago, even though Ref.47 does not refer to Ref.16. Ref.47 does not claim a universal role for thermal vibrations in altering energy differences between various phases. Given the motivation of Ref.47, such a universal conclusion is not warranted. In addition, it would be an exaggeration to generalize to all systems results obtained from a single computational study. However, such a universal claim was made in Ref.16 thirty years ago and is correct as it is based on universal experimental observations mentioned above. *Hence, the role of thermal vibrations must be incorporated in all first principles studies of phase changes.* Since ordering energies, which are in the range 10-100 meV [8], are smaller than energy differences between competing phases, it follows that the role of thermal vibrations must be incorporated in all first principles studies order-disorder transitions. Because the results of Ref.47 can be seen as a confirmation of the prediction made in Ref.16 (that is also valid for order-disorder transitions), it follows that AIMD studies on the role of thermal vibrations in phase transitions will reveal changes in ordering energy at finite temperatures.

It is stressed that the definition of electron-phonon interactions in Ref.15,16 and Ref.28 is slightly different from the conventional definition. In the conventional definition, electron-phonon interactions result only in self-energy effects and AIMD does not incorporate electron-phonon interactions as is does not include self-energy corrections. In the EPI formalism used in Ref.15,16 and Ref.28 electron-phonon interactions result in two correction terms, self energy and DWF corrections. The EPI formalism is a complete theory that includes all effects due to thermal vibrations and the same is explicitly stated in Ref.48. Hence, the full EPI formalism includes the effects that are studied by the AIMD formalism.

Before examining magnetic phase transitions, we briefly summarize the main conclusions of the analysis of alloy phase transitions. The universal experimental observation that thermal vibrations alter total energies by the same order of magnitude as ordering energy implies that the assumption of theoretical models that ordering energy is a constant is highly suspect. Since diffraction data are <u>universally</u> corrected for thermal vibrations, all theoretical models of phase transitions are incorrect and must be corrected to incorporate the role of thermal vibrations. The ES(DWF) formalism for electronic structure calculations is completely compatible with the diffraction theory, due to which, diffraction data <u>universally</u> show that ordering energy is a function of temperature. In contrast, standard references in alloy theory [7-9] assume that ordering energy is independent of temperature, which can now be seen to be incorrect. *<u>A simple criterion is proposed that ordering energy can be assumed to be independent of temperature only if the DWF correction can be ignored in the analysis of diffraction data.</u>*

**5. Magnetic Phase Transitions**

An important question is whether the conclusions from the analysis of alloy phase transitions are of sufficient generality to be valid for magnetic phase transitions as well.

Magnetic phase transitions are readily detected using neutron diffraction techniques [44-46] as ordering of spins gives rise to additional intensity either at new superlattice lines in case of anti-ferromagnetism or at fundamental lines in case of ferromagnetism. Magnetic scattering amplitudes and form factors are reduced by thermal vibrations at finite temperatures and are corrected by a Debye-Waller factor [44-46]. These experimental observations for magnetic ordering are identical to those in binary alloys where ordering is detected by x-ray diffraction. Therefore, the main conclusions from analysis of alloy ordering can be extended to magnetic phase transitions.

In addition, the following argument also suggests a role of thermal vibrations in magnetic phase transitions. Magnetism originates in the exchange interaction involving the wavefunction of the unpaired electron and the total magnetic ordering energy depends crucially on the exchange energy. At finite temperatures, various nuclear configurations are accessed due to thermal vibrations and the electron distribution is in instantaneous equilibrium with them. Assuming that the exchange integral is independent of temperature is equivalent to assuming that the unpaired electron wavefunction is unchanged from the 0 K wavefunction for all configurations that are accessed due to thermal vibrations. It follows that the unpaired electron density and the magnetic form factor are identical at 0 K and finite temperatures. However, this is contradicted by neutron diffraction data where the magnetic form factor has a DWF dependence on temperature. This proves that the assumption that the unpaired electron wavefunction is unchanged by thermal vibrations at finite temperatures is incorrect. Therefore, the thermal-average exchange integral and energy is a function of temperature. *A simple criterion is that the exchange energy can be considered to be a constant only if the DWF can be ignored in the analysis of magnetic neutron diffraction data*.

The ES(DWF) formalism [15, 29-35] allows the determination of band structures of all solids including magnetic solids at finite temperatures. Because this formalism uses temperature dependent core potentials, Eq.7 and Eq.13, the electron wavefunction and exchange integral are temperature dependent due to which the latter must be represented as $J_{ij}(T)$. The role of thermal vibrations in both the ES(DWF) formalism and diffraction theory is incorporated by the DWF. Therefore, if the exchange energy is assumed to be independent of temperature, it implies that the DWF is ignored in the ES(DWF) formalism from which it follows that DWF must be ignored in the analysis of diffraction data as well. That is, *exchange energy can be considered to be a constant only if the DWF can be ignored in the analysis of magnetic neutron diffraction data*.

Most significantly, in the analysis of neutron diffraction data, the role of thermal vibrations is incorporated. The discussion on pages 7-8, including Eq.2 and Eq.3, are correct for all diffraction techniques and are valid for magnetic neutron diffraction as well as seen from standard references [21,22]. As in the case of alloys, illustrated for beta brass, β-CuZn, if the role of thermal vibrations is ignored in analysis of diffraction data, different superlattice lines lead to different magnetic order parameters and it is not possible to speak of an order parameter. Hence, the role of DWF must be incorporated in the analysis of magnetic neutron diffraction data. This shows that the assumption of most models, Eq.1, that incorporating or ignoring the role of thermal vibrations has negligible effect on phase transitions is incorrect. Since DWF is <u>always</u> incorporated in analysis of diffraction data, it follows that the magnetic form factor and the exchange integral are temperature dependent. Therefore, magnetic order parameter obtained from diffraction data can only be compared with theoretical models that also incorporate the role of thermal vibrations, or else the comparison is invalid. *Thus, analysis of magnetic neutron* *diffraction data constrains all theoretical models of magnetic phase transitions to incorporate the role of thermal vibrations.*

In the case of band ferromagnetism in Fe, Co and Ni, the density of states (DOS) at the Fermi level is a very important parameter [27]. Incorporation of the DWF alters the DOS [31] in Cd metal. AIMD simulations on Mo [47] also show that thermal vibrations alter the DOS at the Fermi level. Clearly, theories that seek to explain band ferromagnetism on the basis of DOS obtained from static lattice (0 K) calculations are incorrect. To correctly understand band ferromagnetism in Fe, Co and Ni, the ES(DWF) must be used in to follow the changes in DOS at the Fermi level with temperature. This formalism will also account for the changes in exchange integral that will follow from changes in band structure at high temperatures.

We next discuss the possibility of isotope effect in phase transitions due to the role of thermal vibrations. The isotope effect is predicted to be a universal effect on all properties due to zero-point vibrations [14]. At low temperatures, vibration amplitudes of different isotopes are different leading to different values of all properties for different isotopes [14]. An isotope effect on band gap and EXAFS in Ge has been observed [49-51] below Debye temperature and attributed to differences in electron-phonon interactions for different isotopes. The isotope effect is observed only at low temperatures where displacement is a function of mass and vanishes at high temperatures where the mean-square displacement is independent of mass [52,53].

In the context of phase transitions, diffraction data clearly show that the role of thermal vibrations is important and cannot be ignored. Since different isotopes will have different mean-square displacements, it will lead to different core potentials, Eq.7 and Eq.13 in the ES(DWF) formalism. It follows that wavefunctions, exchange integrals and hence, transition temperatures will be different for different isotopes. Therefore, *an isotope effect on magnetic phase transitions is predicted if $T_c$ is below the Debye temperature, $\Theta_D$.* In alloy phase transitions, $T_c$ is higher than $\Theta_D$ and hence, an isotope effect in alloy phase transitions is highly unlikely.

Recent experimental results [54,55] where an isotope effect has been found in magnetism confirm the above predictions. A change in the Neel temperature $\Delta T_N/T_N \sim 4\%$ and critical magnetic field $\Delta B_C/B_C \sim 4\%$ is reported for different isotopes in Ref.54. The authors of Ref.54 comment that since the material "*is nonmetallic and so Fermi surface effects have no part in the observed isotope effect*" and attribute the isotope effect to the changes in vibrational amplitudes due to different isotopes. The authors of Ref.55 state "*We propose that the smaller J' for deuterated NDMAP is caused by the smaller zero-point motion of deuterons leading to less overlap of the electronic wavefunctions in the exchange paths involving hydrogens*". These authors attribute the isotope effect to the differences in vibration amplitudes for different isotopes and clearly suggest that the exchange integral is a function of vibration amplitude and must be written as $J(\overline{u^2})$.

However, semiconductors are non-metallic as well and as Ref.49-51 show, self energy effects and lattice expansion effects cannot be ignored. Therefore, the conclusion in Ref.54,55 that only the DWF contribution to the isotope effect is significant should be considered as a preliminary conclusion and more experiments are necessary to separate out the contributions of different phenomena.

The most significant conclusion from the isotope effect observed in Ref.54,55 is that it *cannot be explained by Eq.1, i.e. by any of the existing models of magnetism such as Ising, Heisenberg, etc. since they do not incorporate the role of thermal vibrations on exchange energies.* The authors clearly state that the result can only be explained if the exchange integral is a function of vibration amplitude, $J(\overline{u^2})$. The isotope effect observed in Ref.54,55 is due to the same universal phenomena of zero-point vibrations that lead to isotope effect in semiconductors [49-51] and therefore <u>universal</u> conclusions can be drawn that all models must incorporate the role of thermal vibrations.

We can also see that the transition temperatures for two models, one which incorporates the role of thermal vibrations and the other which doesn't will be very different. This is seen more clearly from the following argument. The isotope effect is due to the <u>relatively small</u> differences, $\Delta\left(\overline{u_1^2} - \overline{u_2^2}\right)$, in zero point vibrational amplitudes for different isotopes. It follows that the differences in wavefunctions and exchange integrals, $J(\overline{u_1^2}) - J(\overline{u_2^2})$, will also be <u>relatively small</u>. This <u>relatively small</u> difference in exchange integrals results in different transition temperatures for different isotopes. In contrast, the differences in vibrational amplitudes between two models where one accounts for vibrational amplitude and the other does not (as it assumes a static lattice) is $\left(\overline{u^2}\right)$, which will be <u>much larger</u>

than $\Delta\left(\overline{u_1^2} - \overline{u_2^2}\right)$. Clearly, the difference in wavefunctions and exchange integrals, $J(\overline{u^2}) - J(0)$, between two such models will be much greater than the small difference, $J(\overline{u_1^2}) - J(\overline{u_2^2})$, between two isotopes that leads to an observable isotope effect. Hence, the differences in critical temperatures, $\Delta T_c/T_c$ between such models is likely to be much larger compared to $\Delta T_N/T_N$ observed due to different isotopes. To obtain correct results, vibration amplitudes must be incorporated in determining wavefunctions and exchange integrals must be represented as $J(\overline{u^2})$ in all theories and models.

## 6. Universal role of diffraction data in constraining theoretical models

On the basis of the observed isotope effect in Ref.54,55, the authors of Ref.54 state *"The similarity between the proposed mechanisms suggests that it* (an isotope effect) *may be a common effect in low dimensional Heisenberg anti-ferromagnets"*. But, from an analysis of magnetic neutron diffraction data, we reached the <u>universal</u> conclusion that an isotope effect can be predicted for all magnetic phase transitions, and not just low dimensional Heisenberg anti-ferromagnets, provided the transition temperature is lower than the Debye temperature. Both, analysis of magnetic neutron diffraction data and the isotope effect observed in Ref.54,55, imply that the exchange energy is a function of temperature due to the role of thermal vibrations. Therefore, in the case of magnetic phase transitions, two different experimental results, diffraction data and isotope effect, constrain theoretical models to incorporate the role of thermal vibrations. In the case of alloy phase transitions, two different experimental results, diffraction data and the fact that the effect of thermal vibrations on ordering energy is of the same order of magnitude as the ordering energy itself, constrain theoretical models to incorporate the role of thermal vibrations. In both alloy and magnetic phase transitions, the role of diffraction data is common. As discussed in detail earlier, the fact that both the ES(DWF) formalism and diffraction theory incorporate the role of thermal vibrations through a DWF leads to a simple criterion. *A simple criterion that emerges is that ordering energy or exchange energy can be assumed to be a constant only if DWF can be ignored in the analysis of diffraction data.* However, this leads to the unacceptable consequence that different superlattice lines result in different order parameters. Therefore, to be consistent with the analysis of diffraction data, which <u>always</u> incorporates the DWF, ordering energy or exchange energy must be temperature dependent. *Thus, diffraction data constrain <u>all models</u> of alloy and magnetic phase transitions to incorporate the role of thermal vibrations.* It follows that the Heisenberg model, Eq.1, that is representative of theoretical models of alloy and magnetic phase transitions, must be modified to

$$H(T) = -\sum J_{ij}(T) S_i \cdot S_j \qquad (17)$$

As mentioned earlier, all properties are renormalized by thermal vibrations due to which they are temperature dependent [14]. The ES(DWF) formalism [15, 29-35] is the first step to determine band structures at finite temperatures to which self-energy corrections must be added. Because it uses temperature dependent core potentials, Eq.7 and Eq.13, the wavefunction and exchange integral are temperature dependent. Therefore, if Eq.1 is the representation of theoretical models under static lattice conditions, the proposed Eq.17 would be the correct representation at finite temperatures. The question that needs to be addressed is if ignoring the temperature dependence and using Eq.1 (as is currently done) instead of Eq.17 justified in practice. This paper shows that it is completely unjustified as *diffraction data conclusively prove* that using Eq.1 is incorrect and the proposed correct form, Eq.17, must be used. This conclusion is also supported by other experimental evidences, changes in total energy with temperature in case of alloy phase transitions and isotope effect in case of magnetic phase transitions.

It is important to note that analysis of diffraction data allows for both possibilities, inclusion or exclusion of the role of thermal vibrations or DWF. This is unlike, for example, band gap measurements at finite temperatures which includes the role of thermal vibrations and does not allow the possibility of excluding their role. With both options being available, if analysis of diffraction data always incorporates the DWF, it clearly implies that the role of thermal vibrations is intrinsic to phase transitions and hence, Eq.17 is the correct representation of theoretical models. We have discussed the example of beta-brass, *β*-CuZn, to highlight the unacceptable consequences of ignoring the role of thermal vibrations.

A fundamental contradiction that persists till date, starting from the earliest studies in phase transitions, is that theoretical models are based on one set of assumptions that ignore the role of thermal vibrations while analysis of diffraction data is based on another set of assumptions that incorporate the role of thermal vibrations. Both theoretical models and analysis of diffraction data must be based on the same set of assumptions for any comparisons between the two to be valid. It is clear from Ref.14 that thermal vibrations renormalize physical properties due to which wavefunctions and electron energies have a *T* dependence and must be represented as $\psi_{nk}(T)$ and $\varepsilon_{nk}(T)$. The ES(DWF) formalism [15] is a general formalism that is valid at all temperatures that gives temperature dependent wavefunctions, $\psi_{nk}(T)$ and electron energies, $\varepsilon_{nk}(T)$. Therefore, it is clear that the exchange integral (or interaction parameter) must be represented as $J_{ij}(T)$ in its general form. *Assuming the static lattice value, $J_{ij}$, is a deliberate choice that must be accompanied by a proper justification, which has never been done till date.* This study shows that diffraction results do not justify assuming the static lattice value, $J_{ij}$, in theoretical models of phase transitions. Therefore, predictions of the order parameter, magnetization, transition temperature, critical exponents etc. only from models represented by Eq.17 and not Eq.1 can be compared with results obtained from diffraction data.

## 7. Conclusion

All theoretical models (Heisenberg, Ising etc.) assume a negligible role for thermal vibrations in alloy and magnetic phase transitions. Analysis of diffraction data *conclusively proves* that this assumption is incorrect. Theoretical models can ignore the role of thermal vibrations only if the role of Debye-Waller Factor is ignored in the analysis of diffraction data. Diffraction data constrain all theoretical models to incorporate the role of thermal vibrations. This conclusion is also supported by other experimental results such as the effect of thermal vibrations on ordering energy which is of the same order of magnitude as ordering energy near transition temperatures and an isotope effect on magnetic phase transitions. An electron-phonon interaction (EPI) formalism that incorporates the Debye-Waller Factor in electronic structure calculations already exists and must be adopted for a correct understanding of phase transitions as it can account for all the different experimental results mentioned above. The discrepancy between experimental and theoretical ordering energy in $Ni_3V$ is evidence for the role of thermal vibrations in altering ordering energy. The inter-nuclear potential energy term converges if zero point vibrations are incorporated and this method can replace the Ewald sum method. The three dimensional Ising model cannot represent order-disorder transition in beta brass, CuZn. An isotope effect is predicted for magnetic phase transitions if the transition temperature is below Debye temperature. The long range order parameter obtained from diffraction data can only be compared with predictions of models that incorporate the role of thermal vibrations and not otherwise.

We thank an anonymous referee for his comments on the mass independence of DWF at high temperatures and drawing attention to Ref.52. We also thank Prof. P.P. Singh, Dept. of Physics, for his comments and suggestions.

1. W. L. Bragg and E.J. Williams, Proc. Roy. Soc. **A 145**, 699 (1934)
2. F. C. Nix and W. Shockley, Rev. Mod. Phys. **1**, 1 (1938)
3. T. Muto and Y. Takagi, *Solid State Physics – Advances in Research and Applications, F. Seitz and D. Turnbull (Eds),* **1**, 193 (1955)
4. R. Kikuchi, Phys. Rev. **81**, 988 (1951)
5. L. Onsager, Phys. Rev. **65**, 117 (1944)

6. T. R. S. Prasanna, arXiv:cond-mat/0404550

7. F. Ducastelle, *Order and Phase Stability in Alloys* (North-Holland, Amsterdam 1991)

8. D. de Fontaine, *Solid State Physics – Advances in Research and Applications, F. Seitz and D. Turnbull (Eds),* **47**, 33 (1994)

9. A. van de Walle and G. Ceder, Rev. Mod. Phys. **74**, 11 (2002)

10. V. Blum, G. L. W. Hart, M. J. Walorski and A. Zunger, Phys. Rev. B **72**, 165113 (2005)

11. P. Hohenberg and W. Kohn, Phys. Rev. **136**, B864 (1964)

12. W. Kohn and L. J. Sham, Phys. Rev. **140**, A1133 (1965)

13. R.G. Parr and W. Wang, *Density-Functional Theory of Atoms and Molecules,* (Oxford, Oxford 1994)

14. P. B. Allen, Phil. Mag. B **70**, 527 (1994)

15. P.B. Allen and V. Heine, J. Phys. C **9,** 2305 (1976)

16. P.B. Allen and J. C. K. Hui, Z. Phys. B **37**, 33 (1980)

17. B.E. Warren, *X-Ray Diffraction*, (Dover, New York, 1990)

18. D. R. Chipman and C. Walker, Phys. Rev. B **5**, 3823 (1972)

19. O. Rathmann and J. Als-Nielsen, Phys. Rev. B **9**, 3921 (1974)

20. M. De Graef, *Introduction to Conventional Transmission Electron Microscopy* (Cambridge, Cambridge, 2003)

21. G. E. Bacon, *Neutron Diffraction* (Clarendon, Oxford 1962)

22. Y. A. Izyumov and R. P. Ozerov, *Magnetic Neutron Diffraction* (Plenum, New York 1970)

23. C. G. Shull, W. A. Strauser and E. O. Wollan, Phys. Rev. **83** 333 (1951)

24. R. F. Stewart and D. Feil, Acta Cryst. **A36**, 503 (1980)

25. V. G. Tsirelson and R. P. Ozerov, *Electron Density and Bonding in Crystals*, (Institute of Physics, Bristol, 1996)

26. P. Coppens, *X-ray Charge Densities and Chemical Bonding*, (IUCr Texts on Crystallography, Oxford, 1997)

27. W. Jones and N. H. March, *Theoretical Solid State Physics, Vol.1*, (Dover, New York, 1985)

28. P. B. Allen, Phys. Rev. B **18**, 5217 (1978)

29. C. Keffer, T.M. Hayes and A. Bienenstock, Phys. Rev. Lett. **21,** 1676 (1968)

30. J. P. Walter, R.R.L. Zucca, M.L. Cohen and Y.R. Shen, Phys. Rev. Lett. **24,** 102 (1970)

31. R.V. Kasowski, Phys. Rev. **187**, 891 (1969)

32. R.V. Kasowski, Phys. Rev. B **8**, 1378 (1973)

33. T. V. Gorkavenko, S. M. Zubkova, V. A. Makara and L. N. Rusina, Semiconductors, **41**, 886 (2007)

34. T. V. Gorkavenko, S. M. Zubkova, and L. N. Rusina, Semiconductors, **41**, 661 (2007)

35. C. Sternemann, T. Buslaps, A. Shukla, P. Suortti, G. Doring and W. Schulke, Phys. Rev. B. **63** 094301 (2001)

36. J. Ihm, A. Zunger and M. L. Cohen, J. Phys. C **12**, 4409 (1979)

37. W.E. Pickett, Comp. Phys. Rep. **9**, 115 (1989)

38. E. Kaxiras, *Atomic and Electronic Structure of Solids*, (Cambridge, Cambridge 2003)

39. J.C. Slater, *Quantum Theory of Molecules and Solids*, vol. 3, (McGraw-Hill, New York, 1967)

40. D. R. Chipman, J. Appl. Phys. **31**, 2012 (1960)

41. N. W. Ashcroft and N. D. Mermin, *Solid State Physics* (Harcourt, New York 1976)

42. M. Barrachin, A. Finel, R. Caudron, A. Pasturel and A. Francois, Phys. Rev. B. **50** 12980 (1994),

43. O. Libacq, A. Pasturel, D.N. Manh, A. Finel, R. Caudron and M. Barrachin, Phys. Rev. B. **53** 6203 (1996)

44. C. Wolverton and A. Zunger, Phys. Rev. B. **52** 8813 (1995)

45. N. A. Zarkevich and D. D. Johnson, Phys. Rev. Lett. **92**, 255702 (2004)

46. M. H. F. Sluiter and Y. Kawazoe, Phys. Rev. B **71**, 212201 (2008)

47. C. Asker, A. B. Belonoshko, A. S. Mikhaylushkin and I. A. Abrikosov, Phys. Rev. B. **77** 220102(R) (2008)

48. K. Baumann, Phys. Stat. Solidi (B), **63**, K71 (1974)

49. C. Parks, A. K. Ramdas, S. Rodriguez, K. M. Itoh and E. E. Haller, Phys. Rev. B **49**, 14244 (1994)

50. M. Cardona and M. L. W. Thewalt, Rev. Mod. Phys. **77,** 1173 (2005)

51. J. Purans, N. D. Afify, G. Dalba, R. Grisenti, S. De Panfilis, A. Kuzmin, V. I. Ozhogin, F. Rocca, A. Sanson, S. I. Tiutiunnikov and P. Fornasini, Phys. Rev. Lett. **100,** 055901 (2008)

52. B. T. M. Willis and A. W. Pryor, *Thermal Vibrations in Crystallography* (Cambridge, Cambridge, 1975)

53. C. Huiszoon and P. P. M. Groenewegen, Acta Cryst. **A28**, 170 (1972)

54. P. A. Goddard, J. Singleton, C. Maitland, S. J. Blundell, T. Lancaster, P. J. Baker, R. D. McDonald, S. Cox, P. Sengupta, J. L. Manson, K. A. Funk, and J. A. Schlueter, Phys. Rev. B **78**, 052408 (2008)

55. H. Tsujii, Z. Honda, B. Andraka, K. Katsumata and Y. Takano, Phys. Rev. B **71**, 014426 (2005)